\documentclass[twocolumn,showpacs,preprintnumbers,amsmath,amssymb]{revtex4}
\usepackage{graphicx}
\usepackage{dcolumn}
\usepackage{bm}

\begin{document}


\title{The entanglement of Heisenberg chain with next-nearest-neighbor interaction}

\author{Shi-Jian Gu$^{1,2}$}
\altaffiliation{Email: sjgu@zimp.zju.edu.cn\\URL:
http://www.phystar.net/}
\author{Haibin Li$^2$}\author{You-Quan Li$^2$}\author{Hai-Qing Lin$^1$}
\affiliation{$^1$Department of Physics, The Chinese University of
Hong Kong, Hong Kong, China} \affiliation{$^2$Zhejiang Institute
of Modern Physics, Zhejiang University, Hangzhou 310027, P. R.
China}

\begin{abstract}
The features of the concurrences of the nearest-neighbor and the
next-nearest-neighbor sites for one-dimensional Heisenberg model
with the next-nearest-neighbor interaction are studied both at the
ground state and finite temperatures respectively. Both
concurrences are found to exhibit different behaviors at the
ground state, which is clarified from the point of view of the
correlation function. The threshold temperature with respective to
different number of sites and the thermal concurrences of the
system up to 12 sites are studied numerically.
\end{abstract}

\maketitle

\section{Introduction}
Recently, much attention has been focused on the entanglement in
the spin
systems\cite{KMOConnor2001,PZandardi2000,LFSantos2003,XWang2001PLA,XWang2002PLA,YSun03,SJGu03}
and indistinguishable particle systems\cite{PZanardi02} due to the
recent-discovered importance of entanglement in the quantum
theory\cite{Nielsen1,SeeForExample}. These systems typically
includes transverse field Ising model and anisotropic Heisenberg
model. And the studies not only reveal the non-trial behavior of
entanglement in the phenomenon of condensed matter physics, such
as quantum phase transition\cite{AOsterloh2002,GVidal2003}, but
also shed new light on the quantum physics. However, most of the
previous works on the spin chain mainly focused on the model with
the nearest spin exchange interaction. And in most cases, the
entanglement of formation between two spin qubits vanishes unless
the two sites are at most next-nearest
neighbors\cite{AOsterloh2002}. Thus it is interesting to
investigate the problem when other kinds of interaction besides
the nearest-neighbor one exist, such as next-nearest-neighbor
interaction. This is not merely a pure theoretical consideration,
whereas, there does exist some quasi-one-dimensional compounds,
such as CuGeO$_3$\cite{MHase93} and NaV$_2$O$_5$\cite{JWBrayb},
manifesting strong evidence of the presence of such interaction.

In this paper, we study the pairwise entanglement of the
nearest-neighbor sites and of the next-nearest-neighbor sites in a
Heisenberg chain with the next-nearest-neighbor exchange both at
finite temperatures and the ground state. The entanglement of
formation, i.e., the concurrence\cite{WKWootters98} is used to
quantify these two quantities. In the following section, we first
introduce the model, then show that the entanglement of formation
can be calculated either from the ground state energy at $T=0$ or
from the partition function at finite $T$. In section
\ref{sec:ground}, we study the properties of the entanglement at
the ground state, and discuss some special cases. Our results show
that the presence of the interaction between the
next-nearest-neighbor sites does not enhance the entanglement
between the nearest-neighbor sites, regardless it is a
ferromagnetic or antiferromagnetic coupling. In section
\ref{sec:thermal}, the thermal concurrences of a 12-site system as
well as the threshold temperature with different size's systems
are investigated. Finally, a brief summary and discussions are
given in section \ref{sec:summary}.

\section{The model formulation}
The Hamiltonian of a Heisenberg chain with the
next-nearest-neighbor interaction and periodic boundary conditions
reads
\begin{eqnarray}
H(J)&=&\sum_{j=1}^{L}\left[ \sigma_j\,\sigma_{j+1} + J
\sigma_j\sigma_{j+2}\right], \nonumber \\
\sigma_1&=&\sigma_{L+1}, \label{eq:Hamiltonian}
\end{eqnarray}
where $L$ is the number of lattice sites, $\sigma_j=(\sigma_j^x,
\sigma_j^y, \sigma_j^z)$ denote Pauli matrices of a spin at $j$th
site, and $J$ is a dimensionless parameter characterizing the
interaction strength between the next-nearest-neighbor sites. The
Hamiltonian is obviously invariant under translation, and
moreover, it has SU(2) symmetry, which manifests the spin
conservation. Thus the reduced density matrix between the
arbitrary two sites takes the form
\begin{eqnarray}
\rho_{jl}=\begin{pmatrix}
  u^+ & 0 & 0 & 0 \\
  0 & w_1 & z & 0 \\
  0 & z^* & w_2 & 0 \\
  0 & 0 & 0 & u^-
\end{pmatrix}\label{eq:reducedmat}
\end{eqnarray}
in the standard basis $|00\rangle, |01\rangle, |10\rangle,
|11\rangle$, and the corresponding concurrence has already been
given\cite{KMOConnor2001}
\begin{eqnarray}
C=2\max\Bigl[\,0, |z|-\sqrt{u^+u^-}\,\Bigr].
\end{eqnarray}
The entities of the reduced density matrix (\ref{eq:reducedmat})
can be calculated from correlation functions $G$, for the present
model, they are
\begin{eqnarray}
&&u^+=u^-=\frac{1}{4}(1+G^{zz}),\nonumber \\
&&z=\frac{1}{4}(G^{xx}+G^{yy}+iG^{xy}-iG^{yx})
\end{eqnarray}
where $G^{\alpha\beta}=\langle\sigma^\alpha\sigma^\beta\rangle$.
Hence the concurrence of arbitrary two sites is given by
\begin{eqnarray}
C=\frac{1}{2}\max\Bigl[\,0,2|G^{zz}|-G^{zz}-1\,\Bigr],
\label{CGzz}
\end{eqnarray}
where the SU(2) symmetry has been taken into account. According to
statistical physics, the correlation function of the
next-nearest-neighbor sites at finite temperatures is
\begin{eqnarray}
G^{zz}_{2}(T)=-\frac{T}{3Z}\frac{\partial Z}{\partial J},
\end{eqnarray}
where $Z$ is the partition function and the subscript 2 denotes
distance between two sites (so does the subscript 1 given below).
At the ground state, by the Hellman-Feynman theorem, we have
\begin{eqnarray}
G^{zz}_{2}\Big|_{T=0}=\frac{1}{3}\frac{d E(J)}{dJ}.
\end{eqnarray}
Then the correlation function of neighboring sites is evaluated as
\begin{eqnarray}
G_1^{zz}=\frac{E}{3 L}-JG^{zz}_2,
\end{eqnarray}
where $E=\langle H\rangle$ is the internal energy of the system.
Hence the key point is to study the two-site correlation function
for the next-nearest neighbors

\section{Ground state concurrence}
\label{sec:ground}

\begin{figure}
\begin{center}
\includegraphics[width=6.8cm]{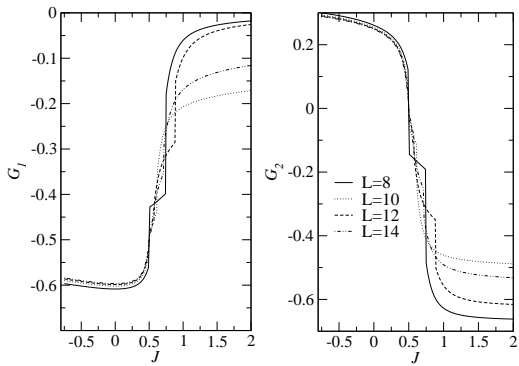}
\caption{The ground state correlation function $G_1$ (left) and
$G_2$ (right) versus $J$ for various sizes. The singularity
at $J=1/2$ arises from the level crossing (or degeneracy).
\label{fig:gzz}}
\end{center}
\end{figure}

\begin{figure}
\begin{center}
\includegraphics[width=6.8cm]{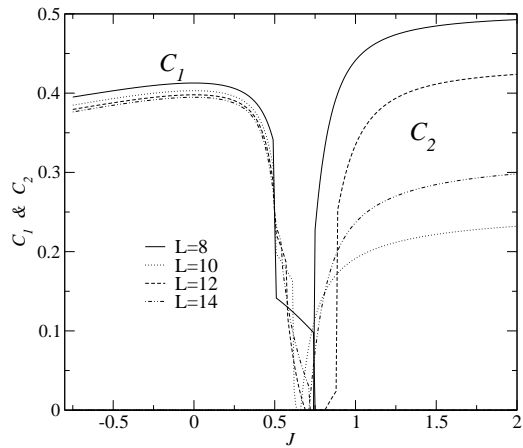}
\caption{The ground concurrence $C_1$ and $C_2$ versus $J$ for
various sizes. Their singularities around $J=1/2$ are caused by
the level crossing (or degeneracy).\label{fig:gcon}}
\end{center}
\end{figure}

As is well known, the exact results of the Hamiltonian
(\ref{eq:Hamiltonian}) for general $J$ have not yet been obtained
except for some special points. At $J=0$, the exact solution for
the ground state and excited states has been well studied by the
Bethe-ansatz method, and the correlation function $G_1^{zz}$ is
simply $E/3L$, thus the thermal concurrence can be expressed in
terms of the internal energy and it equals to 0.386 for the ground
state. When $J=1/2$, the ground state consists of an equal-weight
superposition of the two nearest-neighbor valence bond
state\cite{CKMajumdar69}:
\begin{eqnarray}
&&|\psi_1\rangle=[1,2][3,4]\cdots[L-1,L]\nonumber \\
&&|\psi_2\rangle=[L,1][2,3]\cdots[L-2,L-1]
\end{eqnarray}
where
\begin{eqnarray}
[i,j]=\frac{1}{\sqrt{2}}(|0\rangle_i|1\rangle_j-|1\rangle_i|0\rangle_j).
\end{eqnarray}
Hence the ground state concurrence can be simply written as
\begin{eqnarray}
C=\left(\frac{1}{2}+\frac{1}{2^{L/2}}\right)
\left(2+\frac{(-1)^{L/2}}{2^{L/2-2}}\right)^{-1},
\end{eqnarray}
which becomes 1/4 in the thermodynamic limit.

In general case, however, we need to solve the eigenvalue problem
of the Hamiltonian for finite-size system numerically. It can be
shown that the ground state of the system for $J<0$ is
antiferromagnetic\cite{EHLieb62}, while for $J>0$, many numerical
results suggested that the ground state is antiferromagnetic for
finite chain\cite{HPBader79}. Thus we only need to work in the
invariant subspace spanned by those states with equal number of
down spins and up spins, i.e., $S^z_{\rm total}=0$. For example,
if $L=10$, the dimension of the subspace is 252. This subspace
should include the eigenstate with the lowest eigenvalue of the
system due to the global SU(2) symmetry. The ground state energy
can be obtained by diagonalizing the Hamiltonian in this subspace.

We show the correlation function of the nearest-neighbor sites
$G_1$ and that of the next-nearest-neighbor sites $G_2$ in Fig.
\ref{fig:gzz}, and the corresponding $C_{1(2)}$ in Fig.
\ref{fig:gcon}. From those two figures, we see that $G_1$ reaches
a minimum at $J=0$, which means that the presence of the
interaction $J$ between the next-nearest-neighbor sites does not
enhance the antiferromagnetic correlation between the
nearest-neighbor sites, regardless it is a ferromagnetic or
antiferromagnetic coupling. From this point of view, we deem that
the presence of interaction with the third party generally
suppresses the entanglement between original bi-parties. This fact
obviously leads to a maximum value of the concurrence $C_1$ at
$J=0$. Moreover, in the region of antiferromagnetically frustrated
coupling $J>0$, the concurrence $C_1$ is strongly affected by the
value of $J$, especially around $J=1/2$, as the antiferromagnetic
correlation of $G_1$ is dramatically broken by the frustration
effect. And $C_1$ then will be suppressed down to zero quickly at
a critical point $J_c$ . On the contrary, the correlation function
$G_2$, as well as the concurrence $C_2$, between the
next-nearest-neighbor sites behave completely in a different way.
From the Hamiltonian, we can easily conclude that the frustrated
interaction $J>0$ helps the formation of antiferromagnetic
correlation $G_2$, thus it is an obviously conclusion that $C_2$
is an increasing function of $J$, as shown in Fig. \ref{fig:gcon}.
Moreover, we can also see from the figures that the concurrence as
well as the correlation function are not smoothly continuous for
all value of $J$. We interpret this phenomena as a consequence of
the ground-state level crossing around $J=1/2$ for finite-size
systems.

\section{Thermal concurrence and threshold temperature}
\label{sec:thermal}

\begin{figure}
\begin{center}
\includegraphics[width=6.8cm]{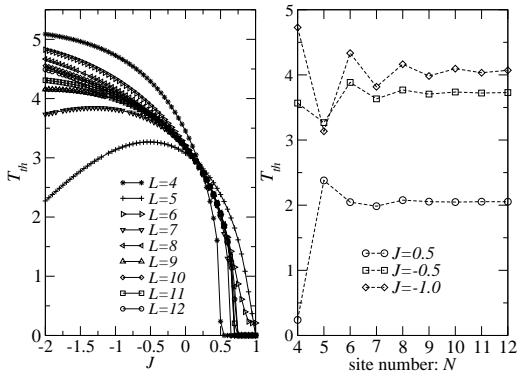}
\caption{Threshold temperature of concurrence of the
nearest-neighbor sites as a function of $J$ for different size.
\label{fig:thtcon}}
\end{center}
\end{figure}

\begin{figure}
\includegraphics[width=6.8cm]{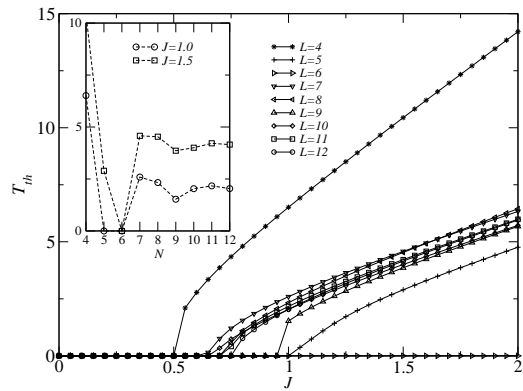}
\caption{Threshold temperature of concurrence between the
next-nearest-neighbor sites as a function of $J$ for different
size. \label{fig:thtcon2}}
\end{figure}

At finite temperatures, the density matrix of the system is the
summation of all states with its Boltzmann weight. Therefore, in
the $T\rightarrow \infty$ limit, the density matrix consists of an
uniform distribution in the state space, which leads to a
vanishing $G^{zz}$. From Eq. (\ref{CGzz}), we then have a zero
entanglement. Thus the thermal fluctuations at high temperature
always suppresses the pairwise entanglement, namely, the
concurrence is a decreasing function of the temperature. So it is
natural to expect that there exists a threshold temperature
$T_{th}$ at which the concurrence vanishes. In this section, we
start with the dependence of threshold temperature on the size of
the system and coupling constant.

We use numerical method to determine the threshold temperature for
the system of $L=4,5,\dots,12$. The results for the concurrence of
the nearest-neighbor sites are presented in Fig. \ref{fig:thtcon},
from which we observe the finite size effect of the threshold
temperature. We find that the threshold temperature converge
quickly as $L$ increases, this is due to the fact that the
concurrence here is expressed in terms of the two-site correlation
function of the nearest-neighbor sites, or from the other point,
simply the internal energy. And for system's energy, a smallish
system, such as $L>12$, can well describe the physical properties
of the thermodynamic system. So we conclude that the system up to
$L=12$ is sufficient to describe the threshold temperature for the
infinite system. We also observe that the system of even (or odd)
number sites manifest different properties. If $J<0$, the
threshold temperature of even-number-site system is larger than
that of odd-number-site system, which is very similar to the
results of traditional isotropic Heisenberg model\cite{XWang02}.
This is because a ferromagnetic $J$ does not frustrated the
tendency of singlet formation between nearest-neighbor sites. If
$J>0$, however, the situation becomes different, merely due to the
frustration. From the left plot of Fig. \ref{fig:thtcon}, we see
that the line of $L=5$ crosses the other lines around $J=0.125$,
so does $L=4$. Consequently, the threshold temperature of the
nearest-neighbor sites is not always a decreasing function of $J$,
the exceptions include $L=5$ and $L=7$. For large system, it is
expected to vanish around $J=0.7$.

\begin{figure}
\begin{center}
\includegraphics[width=7cm]{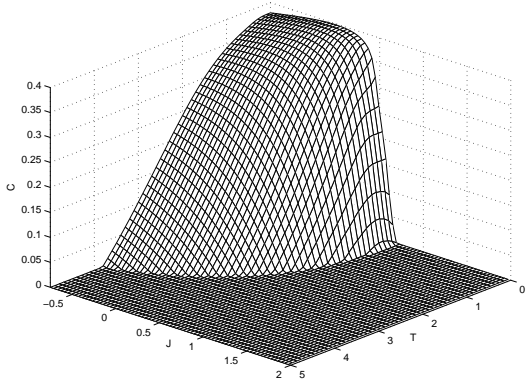}
\caption{Representation of thermal concurrence $C$ between the
nearest-neighbor sites as a function of the temperature and the
next-nearest-neighbor interaction $J$. Here
$L=12$.\label{fig:or12tempcon}}
\end{center}
\end{figure}

\begin{figure}
\begin{center}
\includegraphics[width=7cm]{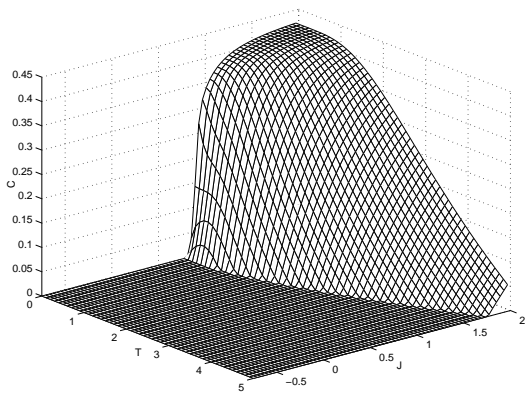}
\caption{Representation of thermal concurrence $C$ between the
next-nearest-neighbor sites as a function of the temperature and
the next-nearest-neighbor interaction $J$. Here
$L=12$.\label{fig:or12tempcon2}}
\end{center}
\end{figure}

In Fig.\ref{fig:thtcon2}, we present our numerical result of the
threshold temperature of the concurrence between the
next-nearest-neighbor sites. We see that for the case of $L=6$,
there is no entanglement at any temperature and $J$. It is because
that if $L=6$, the system can be divided into two parts, including
sites $\{1,3,5\}$ and $\{2,4,6\}$, respectively. For each closed
part, the frustrated interaction does not assist the entanglement
of formation because of the inside-frustration. For other cases,
the threshold temperature exhibits ``down-down-up-up" behavior as
the number of sites increases, which differs from that of the
nearest-neighbor sites completely. And it converges quickly as $L$
exceeds 10.

As we pointed out before, the entanglement of a smallish system
can well represent the behavior of a large system. We show the
thermal concurrence of nearest-neighbor sites as a function of the
temperature and the interaction $J$ in Fig. \ref{fig:or12tempcon}.
Generally, the thermal fluctuation suppresses the pairwise
entanglement. Around $J=0.6$, the concurrence tends to zero at the
ground state, the thermal fluctuation may enhance it. In Fig.
\ref{fig:or12tempcon2}, we show the results of thermal concurrence
of the next-nearest-neighbor sites, from which we can see that
there is no entanglement at any temperature if $J<J_c$. While at
$J>J_c$, the thermal frustration also generally breaks the
entanglement except in a small region of $J\rightarrow J_c$.

\section{Summary and discussion }
\label{sec:summary}

In this paper, we studied the entanglement between the
nearest-neighbor sites as well as the next-nearest-neighbor sites
in a Heisenberg chain with the next-nearest-neighbor interaction
both at the ground state and finite temperatures. We found that
the presence of a frustrated interaction $J>0$ will induce the
entanglement of formation between the next-nearest-neighbor sites.
Thus, the entanglement in spin system stems from sufficiently
large interaction between two sites . However, physically, the
long-range interaction usually decreases as the distance
increases. A typical example is the Haldane-Shastry
model\cite{Haldane88,Shastry88}, which is modelled by
$H=\sum_{nm}J_n\sigma_m\sigma_{m+n}$ with
$J_n=J_0/2\sin^2(n\pi/N)$. Its correlation function
$\langle\sigma_0\sigma_n\rangle$ also decreases rapidly as $n$
increases, which results in a zero concurrence even between the
nearest-neighbor sites. It is therefore believed that the
concurrence is unwonted resources in realistic condensed matter.
We also found that the presence of the interaction with third
party generally suppresses the entanglement between original
bi-parties, regardless it is ferromagnetic or antiferromagnetic.

At finite temperature, we studied the entanglement by considering
the thermal fluctuation. The threshold temperature of the
entanglement was discussed for different system sizes. We found
that $T_{th}$ shows different properties for the system with even
and odd number sites, and it converges quickly when the number of
sites exceeds 10. The thermal fluctuation generally suppresses the
entanglement at finite temperatures except in some region, i.e.,
around $J=0.6$, where it may enhance the entanglement.

This work is supported by the Earmarked Grant for Research from
the Research Grants Council (RGC) of the HKSAR, China (Project
CUHK 4246/01P \& 4037/02P), and NSF China No.10225419 \& No.
90103022.


\begin{thebibliography}{99}
\bibitem{KMOConnor2001}
K. M. O'Connor and W. K. Wootters, Phys. Rev. A {\bf 63}, 052302
(2001).

\bibitem{PZandardi2000}
P. Zanardi, Phys. Rev. A {\bf 65}, 042101(2002).

\bibitem{LFSantos2003}
L.F. Santos, Phys. Rev. A {\bf 67}, 062306 (2003).

\bibitem{XWang2001PLA}
X. Wang, Phys. Lett. A {\bf 281}, 101(2001).

\bibitem{XWang2002PLA}
X. Wang, and P. Zanardi, Phys. Lett. A {\bf 301}, 1(2002).

\bibitem{YSun03}
Y. Sun, Y. Chen, and H. Chen, Phys. Rev. A {\bf 68}, 044301
(2003).

\bibitem{SJGu03}
S. J. Gu, H. Q. Lin, and Y. Q. Li, Phys. Rev. A {\bf 68}, 042330
(2003).

\bibitem{PZanardi02}
P. Zanardi, and X. Wang, J. Phys. A {\bf 35}, 7947 (2002).


\bibitem{Nielsen1}
M. A. Nilesen and I. L. Chuang, {\it Quantum Computation and
Quantum Information} (Cambridge University Press, Cambridge,
England, 2000)

\bibitem{SeeForExample}
See review article by C. H. Bennett and D. P. Divincenzo, Nature
{\bf 404}, 247 (2000).


\bibitem{AOsterloh2002}
A. Osterloh, Luigi Amico, G. Falci and Rosario Fazio, Nature {\bf
416}, 608 (2002).

\bibitem{GVidal2003}
G. Vidal, J. I. Latorre, E. Rico, and A. Kitaev, Phys. Rev. Lett.
{\bf 90}, 227902 (2003).

\bibitem{MHase93}
M. Hase, I. Terasaki, and K. Uchinokura, Phys. Rev. Lett. {\bf
70}, 3651 (1993).

\bibitem{JWBrayb}
J. W. Bray {\it et al.}, in {\it Extended Linear Chain Compounds},
edited by J. S. Miller (Plenum, New Youk, 1993), Vol. 3, pp.
353-415.


\bibitem{WKWootters98}
W. K. Wootters, Phys. Rev. Lett. {\bf 80}, 2245 (1998).


\bibitem{CKMajumdar69}
C. K. Majumdar, J. Phys. C {\bf 3}, 911 (1969).

\bibitem{EHLieb62}
E. Lieb and D. Mattis, J. Math. Phys. {\bf 3}, 749 (1962).

\bibitem{HPBader79}
H. P. Bader and R. Schilling, Phys. Rev. B {\bf 19}, 3556 (1979).

\bibitem{XWang02}
X. Wang, Phys. Rev. A {\bf 66}, 044305 (2002).

\bibitem{Haldane88}
F. D. M. Haldane, Phys. Rev. Lett. {\bf 60}, 635 (1988).

\bibitem{Shastry88}
B. S. Shastry, Phys. Rev. Lett. {\bf 60}, 639 (1988).
\end{thebibliography}
\end{document}